\documentclass[useAMS,usenatbib]{mn2e}
\usepackage{graphics}
\usepackage{graphicx}

\title[SFS models as an alternative to Dark Energy?]{Sudden Future Singularity models as an alternative to Dark Energy?}

\author[H. Ghodsi, M. A. Hendry, M. P. Dabrowski, T. Denkiewicz]{Hoda Ghodsi$^{1}$\thanks{h.ghodsi@astro.gla.ac.uk}, Martin A. Hendry$^{1}$\thanks{martin@astro.gla.ac.uk}, Mariusz P. D\c{a}browski$^{2,3}$\thanks{mpdabfz@wmf.univ.szczecin.pl} and \newauthor Tomasz Denkiewicz$^{2,3}$ \thanks{atomekd@wmf.univ.szczecin.pl}\\
$^{1}$SUPA, School of Physics and Astronomy, University of Glasgow, Glasgow G12 8QQ, UK\\
$^{2}$Institute of Physics, University of Szczecin, Wielkopolska 15, 70-451 Szczecin, Poland\\
$^{3}$Copernicus Center for Interdisciplinary Studies, S\l awkowska 17, 31-016 Krak\'ow, Poland}

\begin{document}



\maketitle

\label{firstpage}

\begin{abstract}
Current observational evidence does not yet exclude the possibility that dark energy could be in the form of phantom energy. A universe consisting of a phantom constituent will be driven toward a drastic end known as the `Big Rip' singularity where all the matter in the universe will be destroyed. Motivated by this possibility, other evolutionary scenarios have been explored by Barrow, including the phenomena which he called Sudden Future Singularities (SFS). In such a model it is possible to have a blow up of the pressure occurring at sometime in the future evolution of the universe while the energy 
density would remain unaffected.  The particular evolution of the scale factor of the universe in this model that results in a singular behaviour of the pressure also admits acceleration in the current era. In this paper we will present the results of our confrontation of one example class of SFS models with the available cosmological data from high redshift supernovae, baryon acoustic oscillations (BAO) and the cosmic microwave background (CMB). We then discuss the viability of the model in question as an alternative to dark energy.
\end{abstract}

\begin{keywords}
cosmology: theory -- cosmological parameters -- cosmology: observations
\end{keywords}

\section{Introduction}

The discovery of the acceleration of the universe \citep{b25}, \citep{b29} resulted in the breaking of the link between geometry and destiny of the universe. It was now the nature of the substance driving this acceleration that would determine the ultimate fate of the universe. Current observations are consistent with a cosmological constant	 as the origin of this acceleration, but they are not yet able to rule out other possibilities such as phantom energy which will drive the universe towards a `Big Rip' singularity. This situation has encouraged the study of other exotic singularities. One such example is the Sudden Future Singularity (SFS) (Type II according to the classification in \citet*{b24}), first proposed by \citet{b4}. Other exotic singularities discovered to date include the Finite Scale Factor (FSF) singularity (Type III \citep{b24}), the Generalised Sudden Future Singularity (GSFS) \citep*{b9}, the Big Separation (Type IV \citep{b8}) and the $w$-singularity \citep{b7}. These singularities may serve as alternatives to dark energy \citep{b8}. Furthermore, the Big Brake is a special kind of a singularity of the SFS type, which can arise in tachyonic models \citep{b16}. In this paper we present the results of our investigation of one example class of models which accommodate an SFS, also proposed by Barrow, by confronting it with the current cosmological observations.

The present work is not the first time that the SFS models have been confronted with cosmological observations. Previously, \citet{b9} performed such a test using the luminosity distance redshift relation applied to supernovae data. Their paper showed that these data were consistent with the SFS model over the redshift range probed by the supernovae, and moreover could permit an SFS that would occur in a suprisingly short time: less than 10 million years from now. In this paper we extend 	the analysis to confront this class of SFS models with some other available cosmological observations and constraints, from the cosmic microwave background radiation, baryon acoustic oscillations and the age of the universe.

The structure of this paper is as follows. In Section 2 we introduce some relevant theory underlying Sudden Future Singularity models. Section 3 gives an account of the cosmological probes we have used to constrain our SFS model, followed by an explanation of the methods we employed to perform these tests. In Section 4 we present the results of our investigations. Finally in Section 5 we give our conclusions.

\section{Sudden Future Singularity Models}

In 2004 \citet{b4} first published the results of his discovery of the existence of Sudden Future Singularities, which arise in the expanding phase of a standard Friedmann universe and violate the dominant energy condition only. This was an intriguing discovery since until then the theoretical search for expanding universes with possible violent ends had identified only Big Rip singularities which violate all the energy conditions \citep{b4}. Consider the Friedmann equations:
\begin{equation}
\rho = \frac{3}{8\pi G} \left(\frac{\dot{a}^2}{a^2}+\frac{kc^2}{a^2}\right),
\end{equation}
\begin{equation}
p = -\frac{c^2}{8\pi G} \left(2\frac{\ddot{a}}{a} + \frac{\dot{a}^2}{a^2} + \frac{kc^2}{a^2}\right),
\end{equation}
where $\rho$ is the energy density, $p$ the pressure, $a$ the scale factor, $k$ the curvature index, $c$ the speed of light, $G$ the gravitational constant and an overdot represents derivative with respect to time. Barrow found that by keeping the scale factor and the Hubble expansion rate finite one will also necessarily maintain a finite density, but that the pressure can still diverge. The divergence of pressure is accompanied by the blow up of the acceleration, leading to a scalar polynomial curvature singularity \citep{b15}. Shortly after their discovery, \citet{b22,b23} showed that these singularities may be avoided (or moderated) when quantum effects are taken into account.

The finiteness of the energy density and the fact that $p \to \infty$ at an SFS means
that the null ($\rho c^2 + p \geq0$), weak ($\rho c^2\geq0$ and $\rho c^2 + p\geq 0$
), strong ($\rho c^2 +p \geq 0$ and $\rho c^2 + 3p \geq 0$) energy conditions are
satisfied but the dominant energy condition ($\rho c^2 \geq 0$, $-\rho c^2\leq p \leq
\rho c^2$) is violated.

We can see that the divergent behaviour of the pressure cannot be linked to the finite
energy density or we will not have an SFS. Hence we need to release the assumption of an
equation of state which imposes a link between these two quantities. It should be noted that an SFS can occur also in inhomogeneous and anisotropic models \citep{b6}, \citep{b4}.

Similar to sudden singularities are the so called `quiescent' singularities which occur in braneworld models \citep{b28}. These are a type of sudden singularities whereby the pressure remains finite alongside the density and the Hubble parameter while higher derivatives of the scale factor diverge. \citet{b1} confronted braneworld models with quiescent singularities with SNe Ia and BAO data and concluded that they would not fit observations. Furthermore the exact same sudden future singularity with the divergent pressure occurs in nonlocal cosmology as found by \citet{b17}. The accelerating universe in this model will lead it towards an SFS rather than a de Sitter type epoch. Koivisto however shows that these singularities may be avoided by a slight modification of the nonlocal model.

Physically sudden future singularities manifest themselves as momentarily infinite peaks of tidal forces, but geodesic completeness is satisfied in our SFS model. Hence Sudden Future Singularities are regarded as weak singularities \citep{b14} which means in-falling observers or detectors would not be destroyed by tidal forces \citep{b8}. The universe will continue its expanding evolution beyond such weak singularities until for example the occurrence of a more serious singularity that is geodesically incomplete like the Big Rip which can end the universe \citep{b9}. 

Barrow constructs an example SFS model where the scale factor takes the form:

\begin{equation}
a(t)=A+Bt^m+C(t_s-t)^n,
\end{equation}
where $A>0$, $B>0$, $m>0$, $C$ and $n>0$ are free constants to be determined. By fixing the zero of time, i.e. setting a(0) = 0, and the time of the singularity, $a(t_s) = a_s$, the scale factor may be written in the equivalent form:

\begin{equation}
a(t) = A+(a_s-A)(\frac{t}{t_s})^m-A(1-\frac{t}{t_s})^n.
\end{equation}
\citet{b9} changed the original parametrisation for the scale factor by using $A=\delta a_s$, to obtain:
\begin{equation}
a(t) = a_s[\delta + (1-\delta)y^m -\delta(1-y)^n] \hspace{0.5cm},\hspace{0.5cm}y=\frac{t}{t_s},
\end{equation}
where $a_s$, $n$, $m$, $\delta$ and $t_s$ are constants to be determined. This way they created a non-standardicity parameter, $\delta$, which, as it tends to zero, recovers the standard Friedmann limit (i.e. a model without an SFS). 

For a pressure derivative singularity to occur we need:
\begin{equation}
r-1 < n < r \hspace{0.5cm},\hspace{0.5cm}  r= \rm{integer}.
\end{equation}
We note that, for $r \geq 3$, all energy conditions are fulfilled. These singularities
are called Generalised Sudden Future Singularities (GSFS) \citep{b9} and may occur in theories with
higher-order curvature corrections \citep{b24}. For an SFS we need $r=2$ which
means that $1<n<2$.

An important requirement is that the asymptotic behaviour of the scale factor close to
the Big Bang singularity follows a simple power law $a_{BB}=y^m$ which will simulate the
behaviour of flat $k=0$ barotropic fluid models with $m=2/3(w+1)$. This will ensure that
all the standard observed characteristics of the early universe such as the CMB, density
perturbations and Big Bang nucleosynthesis are preserved. All the energy conditions are
satisfied if we require $m$ to lie in the range 

\begin{equation}
0<m<1. 
\end{equation}
Furthermore, in accordance with (7), if for a standard dust-dominated universe we take $m=2/3$ our SFS model will reduce to the Einstein-de-Sitter (EdS) universe at early times.

The other parameters of the model that we should consider are: $a_s$, $t_s$ and $\delta$. The parameter $a_s$, which sets the physical size of the universe at the time of the SFS, will cancel out in the equations of the standard cosmological probes we have used to test the model. Thus we do not need to consider this parameter further.

For constraining the time, $t_s$, when an SFS might occur, we can introduce the dimensionless parameter $y_0=t_0/t_s$, where $t_0$ is the current age of the universe. Since the singularity is assumed to be in the future, it follows that $0< y_0 <1$.

The current acceleration obtained in the SFS model, as a result of the particular form adopted for the scale factor, leads to a divergence of the pressure at sometime in the future evolution of the model. \citet{b8} refer to the cause of this late time acceleration in the SFS model as a `pressure-driven dark energy'.

With the parameters $n$, $m$ and $y_0$ having definite values or ranges, it remained to identify a suitable range of investigation for the non-standardicity parameter $\delta$. In \citet{b9} negative values of $\delta$ were associated with acceleration, but we revisited this question in order to check rigorously the range of values of $\delta$ which should be considered, taking into account all relevant physical constraints. In so doing we imposed the established observational facts that both the 
first and second derivatives of the scale factor are currently positive -- i.e. we rejected any combination of SFS model parameters for which $\dot{a}_0 \leq 0$ or $\ddot{a}_0 \leq 0$. In addition we imposed the physical condition that positive and negative redshifts should correspond to past and future events respectively.  Finally we checked the sign of $\dot{a}$ throughout the evolution of the SFS model and rejected parameter combinations that would predict contraction (i.e. $\dot{a} < 0$) at any point in the interval $0 \leq z \leq 1$, on the (conservative) grounds that the expansion of the universe is securely observed from e.g. the Hubble diagram of type Ia supernovae over this range of redshifts.  We did not, however, make any further assumptions about the expansion history of the universe outside of this range; in particular parameter combinations that would predict e.g. future contraction of the universe were not excluded from our analysis.

Thus by fixing $m = 2/3$, as previously discussed, and varying $n$ and $y_0$ over their permitted ranges, we sought to identify those values of $\delta$ which were consistent with the above physical constraints.  It quickly became apparent that this task was not possible analytically.  Therefore we carried out a numerical exploration of the 3-dimensional parameter space $(n,y_0,\delta)$. This then
told us that $\delta$ should not be positive.

\section{Cosmological Constraints and Analysis Methods}
We carried out a comparison between our SFS model and current cosmological observations within a Bayesian framework, computing posterior distributions for the SFS model parameters inferred from each of the cosmological probes considered. From Bayes' theorem we may write

\begin{equation}
p(\Theta | {\rm data}) \propto p({\rm data}| \Theta) p(\Theta),
\end{equation}
where $\Theta$ denotes the parameter(s) of the SFS model and `data' denotes generically the observed data for one of the cosmological probes under consideration. The term on the left hand side of eq. (8) is the posterior distribution for $\Theta$, i.e. it describes our inference about the model parameters in the light of the cosmological data with which we are confronting the SFS model. The first term on the right hand side is the likelihood function which gives the probability of obtaining the data that we actually observed, given a particular set of model parameters. The second term on the right hand side is the prior probability distribution for the model parameters, based on e.g. theoretical considerations and/or previous observations. For each of our cosmological probes we took the likelihood function to be Gaussian in form, i.e.

\begin{equation}
p({\rm data} | \Theta) \propto  \exp ( -\frac{1}{2} \chi^2),
\end{equation}
where the constant of proportionality could be determined from the normalisation conditions that the Likelihood function and posterior probability distribution should integrate to unity. In eq. (9) the quantity $\chi^2$ takes the general form:

\begin{equation}
\chi^2 =  \sum^n\frac{(x_{\rm obs} - x_{\rm pred})^2}{\sigma^2},
\end{equation}
where $x_{\rm obs}$ is the observed value of the relevant cosmological probe, $x_{\rm pred}$ is the value predicted by our SFS model given the parameters $\Theta$, $\sigma$ is the uncertainty (due to measurement error and/or intrinsic variations) in the observed value and $n$ is the number of data points associated with it. We now briefly describe each cosmological probe in more detail.

\subsection{SNe Ia redshift-magnitude relation}
\begin{figure}
 \includegraphics[width=84mm]{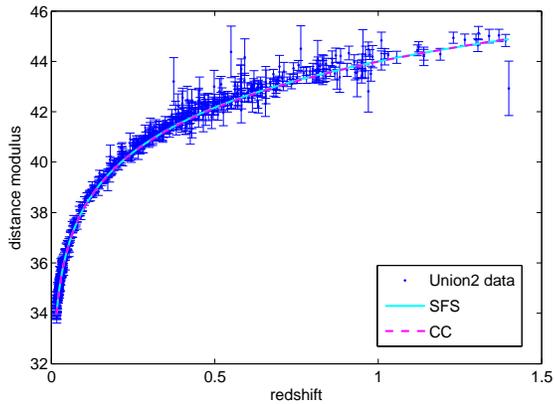}
 \caption{The predicted distance modulus plotted against redshift for the SFS model and Concordance Cosmology standard model (CC). The SFS model prediction is shown as the solid cyan line and is for parameters $n = 1.995$, $\delta = -0.5$, $y_0 = 0.805$. (These are the best-fitting SFS parameters for the Union2 SNe dataset). The concordance model prediction is shown as the dashed pink line and is for parameters $\Omega_m = 0.2725$, $\Omega_\Lambda = 0.7275$ (from the WMAP7 results \citep{b19}). Both sets of calculations assume $H_0 = 74.2$ km s$^{-1}$ Mpc$^{-1}$ from the HST Key Project results \citep{b26}. Also shown are the observed data points (with quoted 1-$\sigma$ errors) for the Union2 SNe dataset. One sees that the SFS model fits the data very well and that the fit is almost indistinguishable from that for the concordance model. (A colour version of this figure is available in the online journal.)}
 \label{fig1}
\end{figure}

We calculated the redshift-magnitude relation for our model to compare it with the observed relation as probed by SNe Ia as standard candles. Although as mentioned in the previous section an SFS could occur regardless of the curvature of the universe, to make our calculations simpler, and also to be consistent with the present observational data, we shall consider only a flat universe which bears an SFS. 
We used 557 SNe from the Union2 dataset as compiled by \citet{b2} which is the largest published SNe Ia sample to date. In our analysis we treated the Hubble constant as a nuisance parameter, marginalising over a range of values using as a prior distribution the most recent results from the Hubble Space Telescope (HST) Key Project \citep{b26}, which assume no underlying cosmology in measuring this parameter.  We describe this marginalisation process in more detail below. For our SNe Ia likelihood function we adopt the $\chi^2$ quantity 

\begin{equation}
\chi^2 = \sum^{n} \frac{(\mu_{\rm obs} - \mu_{\rm pred})^2}{\sigma_{\rm phot}^2 + \sigma_{\rm int}^2},
\end{equation}
where $\mu$ denotes distance modulus, $\sigma_{\rm phot}$ is the measurement uncertainty tabulated for each SN and $\sigma_{\rm int}$ is the scatter intrinsic to every supernova because they are not perfect standard candles. We take $\sigma_{\rm int}$ to be 0.15 as advocated by \citet{b20}.\\
\newline
Fig.~\ref{fig1} shows the distance modulus as a function of redshift for the SFS model and the concordance model for their respective best-fitting parameters and how they compare with the best SNe dataset available. One can see that the SFS model fits the data very well and that it is almost distinguishable from the concordance model over the redshift range probed by the Union2 sample.

\subsection{CMB distance priors: $R$ and $l_a$}
In order to test a model against the cosmic microwave background (CMB) data one would ideally calculate the full angular power spectrum of the temperature anisotropies for the model, and compare it with the observed angular power spectrum. However, since this process is rather computationally intensive and complex, a simpler approach is instead to calculate the distance
scales to which the power spectrum is very sensitive. The positions of the peaks and troughs of the CMB power spectrum, which can be measured precisely, provide a measure of the distance to the decoupling epoch. The distance ratios measured by the CMB are:\\
\newline(1) the angular diameter distance to the last scattering surface (at redshift $z_{\rm CMB}$, which we take to be 1089), $D_A(z_{\rm CMB})$, divided by the sound horizon size at the decoupling epoch, $r_s(z_{\rm CMB})$ which is quantified by the `\emph{acoustic scale}' and is defined by:
\begin{equation}
l_a = (1+z_{\rm CMB}) \frac{\pi D_A(z_{\rm CMB})}{r_s(z_{\rm CMB})},
\end{equation}
where the factor $(1+z_{\rm CMB})$ accounts for the fact that $D_A(z_{\rm CMB})$ is the proper angular diameter distance and we calculate $r_s(z_{\rm CMB})$ following \citet{b31}.\\
\newline
(2) the angular diameter distance to the last scattering surface divided by the Hubble horizon size at the decoupling epoch, which is called the `\emph{shift parameter}', $R$ and is given by \citet{b18}:
\begin{equation}
R(z_{\rm CMB}) = \frac{\sqrt{\Omega_m H_0^2}}{c} (1+z_{\rm CMB}) D_A(z_{\rm CMB}),
\end{equation}
where $\Omega_m$ is the matter density parameter. It was shown by \citet{b31} that using these two parameters together is necessary in order to place tight CMB constraints on the parameters of the model of interest. They found that models with the same parameter $R$ but different values for $l_a$, and vice versa, in general gave different CMB angular power spectra.

For the CMB constraints we therefore computed likelihoods with the following $\chi^2$ expressions for the shift parameter and the acoustic scale respectively:

\begin{equation}
\chi^2_{\rm R}  =  \frac{(R^{\rm obs} - R^{\rm pred})^2}{\sigma_{\rm R}^2},
\end{equation}

\begin{equation}
\chi^2_{l_a}  =  \frac{(l_a^{\rm obs} - l_a^{\rm pred})^2}{\sigma_{l_a}^2}.
\end{equation}

The observed values for these quantities: $R= 1.725 \pm 0.018$ and $l_a= 302.09\pm0.76 $, are obtained from the WMAP7 data and given by \citet{b19}. However, as pointed out by \citet{b13} the shift parameter is not a directly measurable quantity and it is in fact derived from the CMB data assuming a specific cosmological model. Care therefore needs to be taken when using this quantity as a cosmological constraint. 

The value for the shift parameter quoted above, as \citet{b18} explains, has been derived assuming a standard FLRW universe with matter, radiation, curvature and dark energy components. The SFS model we are considering is also a standard Friedmann model but it assumes no explicit dark energy component; instead cosmic acceleration is driven by the divergence of pressure resulting from the particular form of scale factor adopted in the model. Matter is also permitted in the SFS and in fact, since we require our model to reduce to the EdS case at early times, we adopt the same matter content as that in the concordance model. Concerning the radiation and curvature components, we follow \citet{b18} and ignore these in the shift parameter calculation.

Turning to the dark energy component, here we followed the approach of \citet{b12} and expressed our SFS model in a form equivalent to an evolving dark energy model by computing its effective equation of state, $w_{\rm eff}(z)$. (See the Appendix for a short derivation of the expression for $w_{\rm eff}$).

\begin{figure}
 \includegraphics[width=84mm]{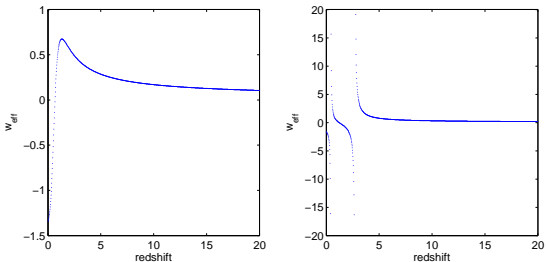}
 \caption{The evolution of $w_{\rm eff}$ as a function of redshift in the SFS model, for two representative sets of model parameters.  Both plots show generic behaviour as $z \rightarrow 0$ and $z \rightarrow \infty$ which is in good agreement with current observations, although plot (b) also shows `fake' singularities due to the parametrisation of $w_{\rm eff}$.}
 \label{fig2}
\end{figure}

We then looked at the evolution of this effective $w$ index to see how it compared with the observed behaviour. Fig.~\ref{fig2} shows the evolution of the effective equation of state over the redshift range $0 < z < 20$, for two representative sets of SFS model parameters. In both cases we see the the same general features: $w_{\rm eff} \simeq -1$ as $z \rightarrow 0$ and $w_{\rm eff} \rightarrow 0$ for large $z$. These limiting behaviours are in good agreement with current observations. Note, however, that
in plot (b) the particular SFS model parameters result in the divergence of $w_{\rm eff}$  at certain redshifts, which is caused by the denominator in the expression on the right hand side of eq. A.5 tending to zero at these redshifts. This behaviour is discussed in \citet{b27} where they make the case that in models where dark energy is not treated as an explicit fluid or a field, the equation of state cannot be used as a fundamental quantity and indeed an effective equation of state may display unusual properties like singularities. Very similar divergent behaviour was found e.g. by \citet{b30} in computing an effective equation of state for their evolving $\Lambda$ model; indeed those authors refer explicitly to each divergent feature as a `fake singularity, which is nothing but an artefact of the EOS parametrisation'. In our case too the divergence of $w_{\rm eff}$ is not seen as an indication of a fundamental physical problem with the model.  Nevertheless the general similarity of the limiting behaviour of $w_{\rm eff}$ to that of the concordance model gives us confidence, following \citet{b13}, that it remains appropriate to use the `observed' value of the shift  parameter when investigating our SFS model.

\subsection{BAO distance parameter, $A$}
Baryon acoustic oscillations which originate from the excitation of sound waves in the early universe photon-baryon plasma through cosmological perturbations are useful distance indicators at the current epoch. We can use this to constrain very well the following quantity, termed as the `\emph{distance parameter}', using the current data:
\begin{equation}
A = \frac{\sqrt{\Omega_m}}{E({z_{\rm BAO}})^{1/3}} \left[\frac{1}{{z_{\rm BAO}}} \int_0^{z_{\rm BAO}} \frac{dz}{E(z)}\right] ^ {2/3}.
\end{equation}
The function $E(z)$ is defined as: $E(z) = \frac{H(z)}{H_0}$, where $H(z)$ is the Hubble expansion rate. $z_{\rm BAO}$ is the effective redshift of the galaxy sample used to measure the distance parameter and it is 0.35 for the SDSS galaxy sample. For the BAO distance parameter constraint we adopt:
\begin{equation}
\chi_{\rm BAO}^2 = \frac{(A_{\rm obs} - A_{\rm pred})^2}{\sigma_{\rm A}^2}.
\end{equation}
Here the observed value for the distance parameter has been taken from the latest SDSS results which is: $A=0.469 (n/0.98)^{-0.35} \pm 0.017$ \citep{b11}. We used the value of $n=0.963$ for the spectral index from WMAP7 results \citep{b19}.

We find that the BAO distance parameter is not immune to the model dependency issue in its derivation either. This issue is considered in detail in e.g. \citet{b5} who, following \citet{b10}, identify two (implicitly assumed) conditions which should be valid in order that the BAO distance parameter is applicable. For the model in question firstly the evolution of matter density perturbations during the matter dominated era must be similar to the Concordance Cosmology case, and secondly the comoving distance to the horizon at the epoch of matter-radiation equality should scale inversely with $\Omega_mH_0^2$. While these conditions are {\em not\/} met for the Carneiro et al. model of a time varying cosmological constant, they are met in our case.

\subsection{Age of the universe, $t_0$}
Using the standard Friedmann equation for calculating the age of the universe we have:
\begin{equation}
t_0 = D_{\rm H} \int_0^\infty \frac{dz}{(1+z)E(z)},
\end{equation}
where $D_H$ is the Hubble distance and $E(z)$ is defined as before. In our likelihood function for this constraint we used the $\chi^2$ expression:
\begin{equation}
\chi_{\rm age}^2 = \frac{(t_{0}^{\rm obs} - t_{0}^{\rm pred})^2}{\sigma_{\rm age}^2} .
\end{equation}
where $t_0^{\rm pred}$ is the age of the universe predicted in our SFS model, as computed from eq. (18). For $t_0^{\rm obs}$, the observed age of the universe, we followed e.g. \citet*{b3} and used the best current estimate derived from various astrophysical probes, including globular cluster ages, that have been determined without assuming a particular cosmological model. We adopt $t_0=12.6^{+3.4}_{-2.2}$ Gyr with an asymmetric error distribution corresponding to 95 per cent upper and lower confidence limits, as reported in \citet{b21}. 

\subsection{Hubble Constant, $H_0$}

The current value of the Hubble constant is the final constraint which we employed in our analysis.  Although the Hubble constant already features indirectly in the SNe Ia and age constraints, since it enters into our calculations of the predicted values for those two probes, we considered it useful also to compare directly the observed value of $H_0$ with its predicted value calculated for our SFS model.  The latter is simply given by:

\begin{equation}
H_0^{\rm pred} = \left( \frac{\dot{a}}{a} \right)_0 = \left( \frac{m(1-\delta)y^{(m-1)}+n\delta(1-y_0)^{(1-n)}}{\delta + (1-\delta)y_0^m-\delta(1-y_0)^n} \right).
\end{equation}
We thus computed a likelihood function for this constraint using the $\chi^2$ quantity

\begin{equation}
\chi_{\rm H_0}^2 = \frac{(H_0^{\rm obs} - H_0^{\rm pred})^2}{\sigma_{\rm H_0}^2},
\end{equation}
where $H_0^{\rm obs} = 74.2 \pm 3.6$ $\rm km$ $\rm s^{-1} \rm Mpc^{-1}$, i.e. the (cosmology-independent) HST Key project value discussed earlier.

\section{results}
            
In this section we present posterior distributions for our SFS model parameters inferred from each of the cosmological probes described in Section 3.  For ease of presentation we have computed a series of
`slices' through the 2-dimensional conditional distribution of the parameters $n$ and $y_0$ at given values of the non-standardicity parameter $\delta$.  We adopted uniform priors for $n$ and $y_0$ over
the intervals $1 < n < 2$ and $0 < y_0 < 1$ respectively, on the theoretical grounds discussed in Section 2.  Thus, for these uniform priors, the mode of the posterior distribution function, for each
`slice' in $\delta$ was coincident with the maximum of the conditional likelihood function.
\\
\newline
For the case of the SNe Ia data, as noted in Section 3 we computed the posterior distribution after first marginalising over the Hubble constant. It follows straightforwardly from Bayes' theorem that:

\begin{equation}
p(n,y_0 | \delta)  \quad = \quad \int p(n,y_0 | \delta) p(H_0) dH_0,
\end{equation}
where $p(H_0)$ represents our prior information on the Hubble constant. Thus for the SNe Ia case we computed $p(n,y_0 | \delta)$ following eq. (22), adopting a Gaussian prior for $H_0$ with a mean value of 74.2 and a standard deviation of 3.6, in accordance with the results of the HST Key Project.
\\
\newline
We computed posterior distributions adopting both a regular grid of $(n,y_0)$ values over the 2-dimensional parameter space and a simple MCMC-based approach using the Metropolis algorithm. Both methods gave very similar results, although the latter approach is more practical when exploring more general parameter spaces with larger numbers of parameters. We will consider such cases further in a subsequent paper.

\begin{figure*}
 \includegraphics[width=170mm]{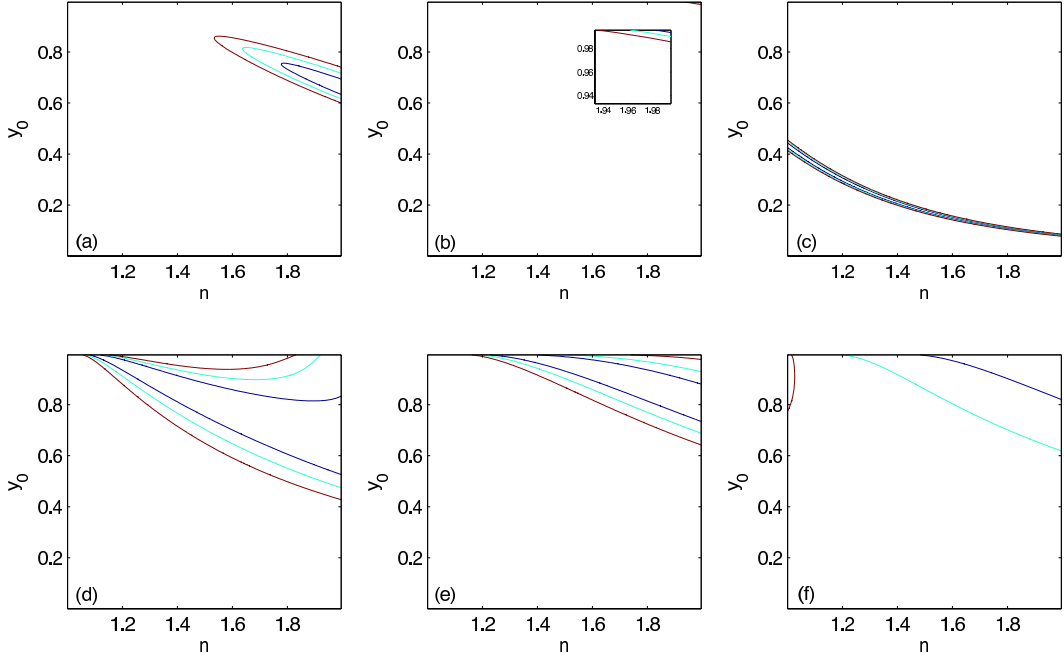}
 \caption{The contours show 68 per cent, 95 per cent and 99 per cent credible regions in the ($n$,$y_0$) parameter space for a fixed value of $\delta=-0.7$. The plots labelled (a), (b), (c), (d), (e) and (f) correspond to the contours calculated using the probes, SNe Ia redshift-magnitude relation, CMB shift parameter, $R$, CMB acoustic scale, $l_a$, BAO distance parameter, $A$, the Hubble constant and the age of the universe respectively.}
 \label{fig3}
\end{figure*}

\begin{figure*}
 \includegraphics[width=170mm]{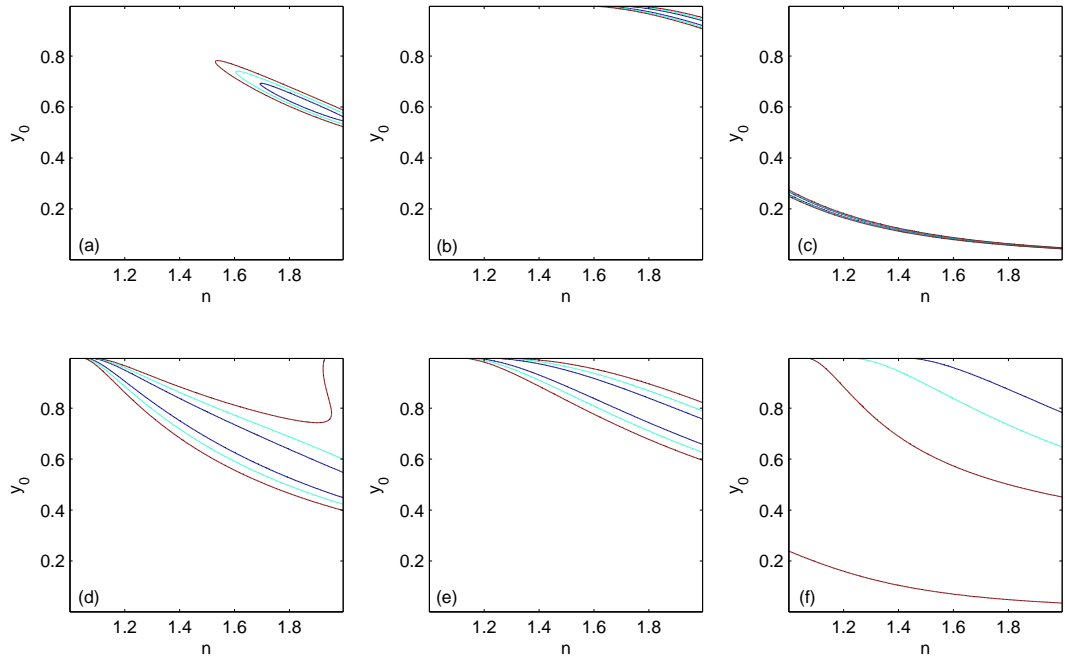}
 \caption{The contours show 68 per cent, 95 per cent and 99 per cent credible regions in the ($n$,$y_0$) parameter space for a fixed value of $\delta=-1$. The labels are as described in Fig.~\ref{fig3}.}
 \label{fig4}
\end{figure*}

\begin{figure*}
 \includegraphics[width=170mm]{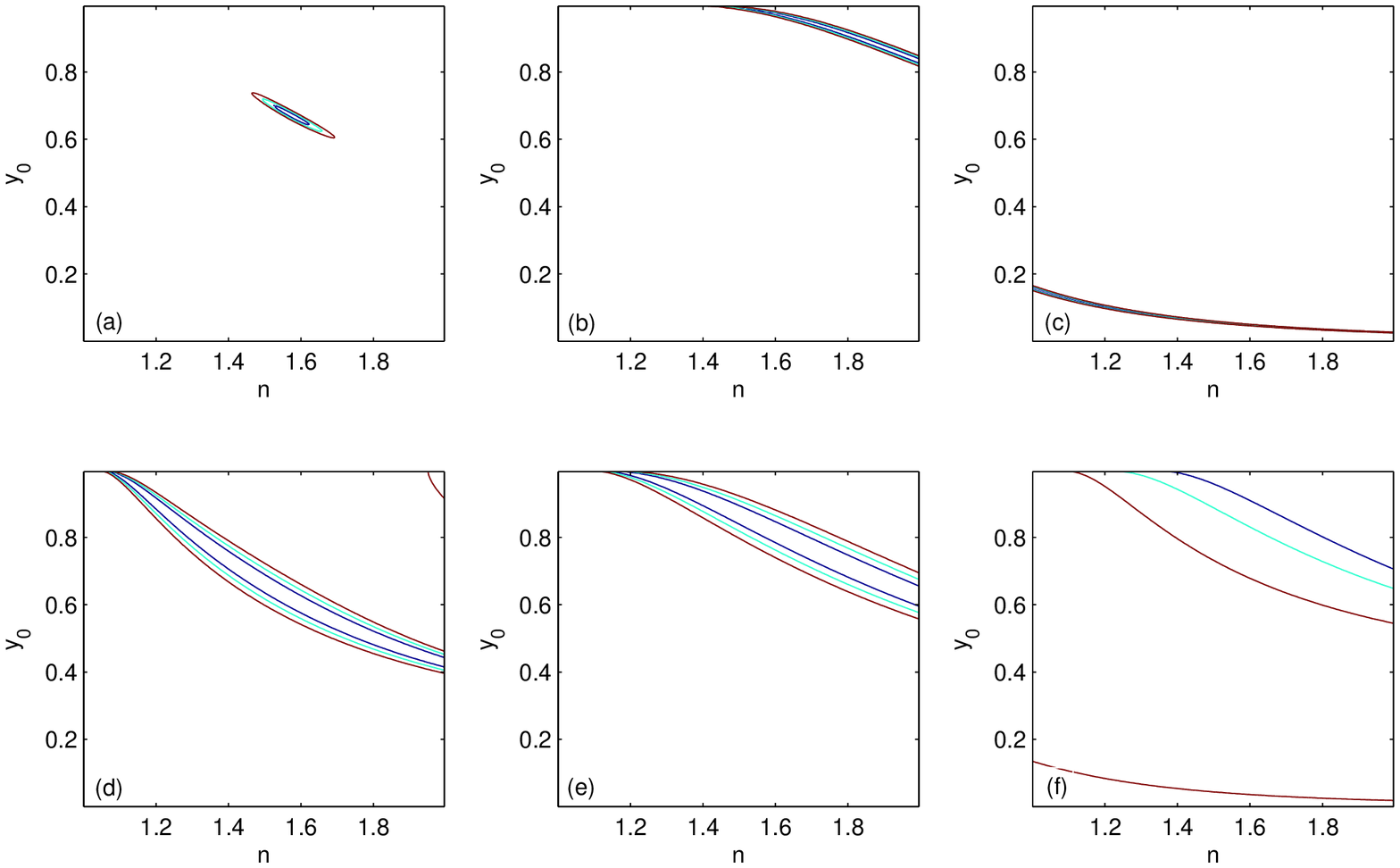}
 \caption{The contours show 68 per cent, 95 per cent and 99 per cent credible regions in the ($n$,$y_0$) parameter space for a fixed value of $\delta=-1.5$. The labels are as described in Fig.~\ref{fig3}.}
 \label{fig5}
\end{figure*}

Figs. \ref{fig3} - \ref{fig5} present contour plots showing the Bayesian credible regions (evaluated at the 68 per cent, 95 per cent and 99 per cent level) for $n$ and $y_0$, computed for each of our cosmological probes at a series of fixed values of $\delta$. In each figure we show separately contours for each of the six cosmological probes we have considered, labelled (a) -- (f) respectively. These probes are: (a) the SNe Ia redshift-magnitude relation; (b) the CMB shift parameter, $R$; (c) the CMB acoustic scale, $l_a$; (d) the BAO distance parameter, $A$; (e) the present-day value of the Hubble constant, $H_0$; (f) the present age of the universe, $t_0$.  Fig.~\ref{fig3} shows results for $\delta = -0.7$. For larger (i.e. less negative) values of $\delta$ the posterior distributions for most of the probes were qualitatively similar; significantly, however, the predicted value of the shift parameter was such that no credible region for the SFS model parameters was found at the 99 per cent level. We return to this point below.

One can clearly see from Fig.~\ref{fig3} that there is no part of the $(n,y_0)$ plane where the credible regions for the six cosmological probes overlap, for this value of $\delta$. There is substantial overlap between the SNe Ia and BAO contours; this is not surprising given that both probes are sensitive to the $E(z)$ function over a similar range of redshifts. 

The age of the universe constraint (f) is rather weak, reflecting the large uncertainty on the observed value, and only a small region of the $(n,y_0)$ plane, on the upper left of Fig.~\ref{fig3}(f), is excluded at the 95 per cent level; these parameter values are already strongly excluded by all other probes.

In Fig.~\ref{fig3}(b) we have magnified the shift parameter contours for greater clarity. Note, however, that the minimum $\chi^2$ value for the shift parameter in this plot is already 21.35 (corresponding to a predicted value of $R=1.641$) which is unacceptably large -- i.e. the likelihood function (and hence the posterior probability) for the SFS model parameters is everywhere vanishingly small for the shift parameter at this value of $\delta$. Similar behaviour was observed for larger values of $\delta$; indeed the minimum $\chi^2$ value becomes progressively higher as $\delta$ increases. Hence we do not show contour plots for $\delta > -0.7$.

It is interesting to note from panels (b) and (c) in Fig.~\ref{fig3} that the two CMB constraints $R$ and $l_a$ produce credible regions which do not in fact show {\em any\/} overlap at the 99 per cent level. This illustrates the importance of the point made by Wang and Mukherjee, as noted earlier, that one should use {\em both\/} CMB probes in order to better constrain (or indeed, as is the case here, to reject) a given model, since these probes are sensitive to the model parameters in different ways. Moreover we also see that the credible regions for $R$ and $l_a$ do not share a common overlap with any of the other four probes either. Hence a fit to the SFS model parameters is strongly excluded for this value of $\delta$.

Figs. \ref{fig4} and \ref{fig5} present credible regions for two further values of $\delta$, in order to illustrate the pattern of behaviour exhibited by the various probes as $\delta$ decreases. Firstly Fig.~\ref{fig4} shows results for $\delta=-1$. We can see that, as before, there appears to be overlap between the credible regions for SNe, BAO distance parameter, Hubble constant and age of the universe; indeed the contours for these probes have shifted only slightly from their position in Fig.~\ref{fig3}. However all four show no overlap at the 99 per cent level with either of the CMB probes, which in turn continue to show no overlap with each other. 

\begin{figure*}
 \includegraphics[width=170mm]{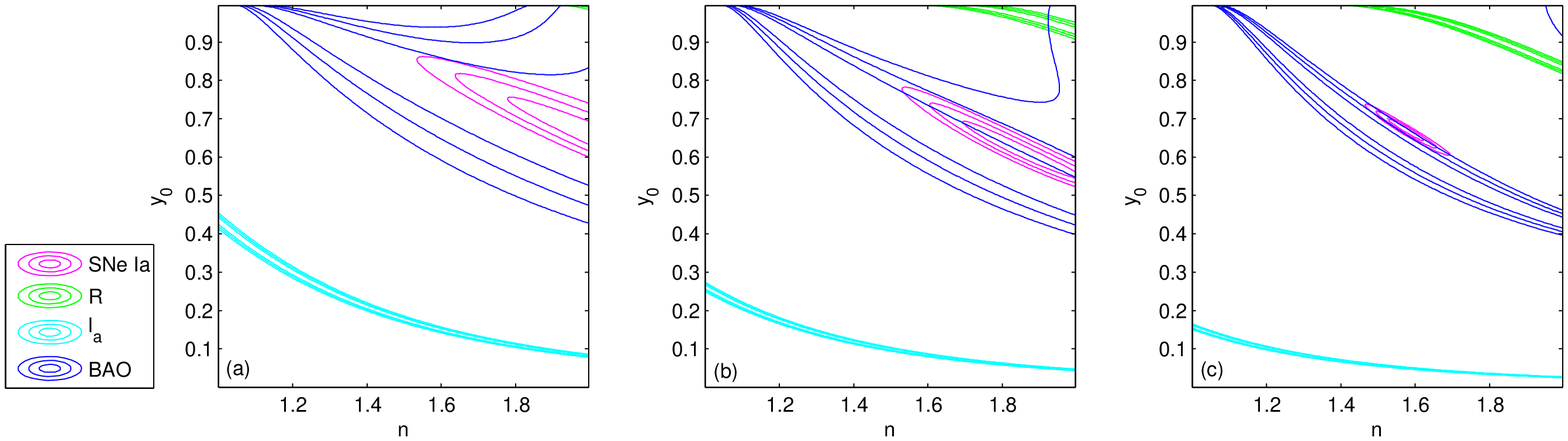}
 \caption{The panels show a superposition of the credible regions in the $(n,y_0)$ parameter space derived from the SNe Ia, CMB shift parameter, $R$, CMB acoustic scale, $l_a$ and BAO distance parameter, $A$ data, for the values of $\delta = -0.7$ (a), $\delta = -1$ (b) and $\delta = -1.5$ (c), as previously illustrated in Figs.~\ref{fig3} - \ref{fig5} respectively. Note that in each panel there is no part of the parameter space where all credible regions overlap. (A colour version of this figure is available in the online journal).}
 \label{fig6}
\end{figure*}

Finally, Fig.~\ref{fig5} shows the credible regions for $\delta=-1.5$. Here we can see that the contours follow more or less the same shapes as in the previous figures. There is still no overlap between $R$ and $l_a$, and neither CMB probe overlaps with any of the other probes, again at the 99 per cent level. Note, moreover, that while the general appearance of the credible regions is similar to Fig.~\ref{fig4}, the contours for the acoustic scale, $l_a$, in Fig.~\ref{fig5}(c) have shifted slightly downwards and appear to be approaching the lower limit of $y_0$. As $\delta$ is decreased further this trend continues and hence no fit is obtained. To emphasize that indeed the contours display no overlap with one another, we show in Fig.~\ref{fig6} the most important contours of SNe Ia, CMB shift parameter, $R$, CMB acoustic scale, $l_a$ and BAO distance parameter superimposed for values of $\delta$ previously considered in Figs. \ref{fig3} - \ref{fig5}. One can see clearly in these figures that while the SNe Ia data are consistent with an SFS occurring in the very near future, as shown by \citet{b9}, the same SFS parameters would not give a fit to the CMB data.

\begin{figure}
 \includegraphics[width=84mm]{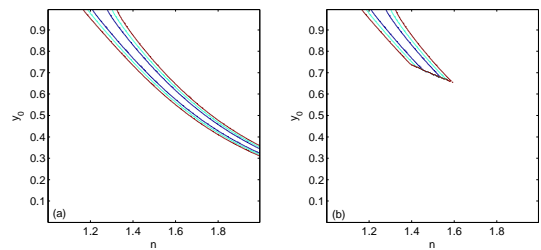}
 \caption{The impact of the imposition of the physical conditions discussed in Section 2 is shown here for the $l_a$ contours. The plot labelled (a) shows these contours not considering the physical conditions while in plot (b) these conditions are included.}
 \label{fig7}
\end{figure}

Note that for the purpose of straightforward illustration the contour plots presented in Figs.~\ref{fig3} - \ref{fig5} have been calculated first without the imposition of the physical constraints discussed in Section 2. A typical example of the application of these conditions, and their impact on the credible regions, is shown in Fig.~\ref{fig7} for the acoustic scale, $l_a$. One can see in Fig.~\ref{fig7}(b) how the contours are abruptly cut off, in this case by the inclusion of the condition that the universe currently be accelerating.  In other words, in the region enclosed by the contours in the right hand part of Fig.~\ref{fig7}(a), e.g. for $y_0 < 0.6$ and $n > 1.5$, our SFS model predicts a value of $l_a$ which is in satisfactory agreement with observations, but for a universe which is not currently accelerating. The other cosmological probes such as the age of the universe and the BAO distance parameter are similarly affected, although much less severely, for certain values of $\delta$. However, since it is already the case that we find no significant overlap between the credible regions over the entire parameter space {\em without\/} applying these additional physical constraints, we will not present any further plots that do include them. 

\section{summary and conclusions}
 
In this paper we have investigated one class of Sudden Future Singularity models proposed by Barrow, by confronting it with the currently available observational data. After introducing the theory behind the model and explaining its characteristics in Section 1, we reported on the cosmological probes we used in testing our model in Section 2, taking care throughout to consider thoroughly the applicability of each probe to the SFS model under study. We then presented the results of our investigations in Section 4.  Specifically we found that the SFS model was not compatible with all of the current cosmological  observations considered:  while good agreement could be found with e.g. the Union2 SNe data and BAO distance parameter at low redshift, the same model parameters predicted an acoustic scale and a shift parameter for the CMB which were in strong disagreement with WMAP7 observations.

We have presented our results as a series of conditional `slices' through the $(n,y_0)$ parameter space at fixed values of $\delta$. We note that the rejection of the SFS model does {\em not\/} depend on our choice of 2-dimensional parameter space on which to conduct our analysis.  The choice of the $(n,y_0)$ space was convenient because of the ranges of these two parameters, but the absence of any region which gave a fit to the model is a robust result that can be extended to the full 3-dimensional space of ($n,y_0,\delta$). In other words, by showing that there is no fit to the data in conditional 2-dimensional spaces we have therefore shown that there is no fit to the data in the full 3-dimensional parameter space.

In conclusion we note that our results are for a fixed value of $m = 2/3$, which ensures that our SFS model reduces to the EdS universe at early times. In a forthcoming paper (Dabrowski, Denkiewicz, Ghodsi \& Hendry in preparation) we will relax this assumption and consider SFS models in which the parameter $m$ is allowed to vary, by confronting these models with the same cosmological observations which we have considered here.

\section*{Acknowledgments}

HG would like to thank Dr Arman Shafieloo for his useful comments and suggestions.

\appendix

\section{Derivation of $w_{\rm eff}$}

In this section we present a short derivation of the expression for $w_{\rm eff}$ discussed in Section 3.2. A dark energy component with equation of state parameter $w(a)$, will correspond to a density that varies with the scale factor according to:

\begin{equation}
\rho(a)=\rho_0 \exp \left( -3\int_1^a \frac{da}{a}[1+w(a)]\right).
\end{equation}

The standard Friedmann equation for a flat model will therefore take the form:

\begin{equation}
\left(\frac{H(a)}{H_0} \right)^2=\frac{\Omega_m}{a^3}+(1-\Omega_m) \exp\left( -3\int_1^a \frac{da}{a}[1+w(a)]\right).
\end{equation}

Re-writing the above using $a = (1+z)^{-1}$ we have:

\begin{equation}
\left(\frac{H(z)}{H_0} \right)^2=\Omega_m (1+z)^3+(1-\Omega_m) \exp\left( 3\int_0^z \frac{1+w(z)d(z)}{1+z}\right),
\end{equation}

which means:

\begin{equation}
\int_0^z \frac{1+w(z)d(z)}{1+z} = \frac{1}{3} \ln \left(\frac{E(z)^2-\Omega_m (1+z)^3}{1-\Omega_m}\right),
\end{equation}
where $E(z)=H(z)/H_0$. Now, differentiating both sides gives, where $'\equiv \frac{d}{dz}$:

\begin{equation}
w(z) = -1 + (1+z) \left( \frac{2/3E(z)E'(z)-3\Omega_m(1+z)^2}{E(z)^2 - \Omega_m(1+z)^3}\right).
\end{equation}

\bsp

\label{lastpage}


\begin{thebibliography}{99}

\bibitem[\protect\citeauthoryear{Alam \& Sahni}{2006}]{b1} Alam U., Sahni V., 2006, Phys. Rev. D., 73, 084024
\bibitem[\protect\citeauthoryear{Amanullah et al.}{2010}]{b2} Amanullah R., et al., 2010, ApJ, 716, 712
\bibitem[\protect\citeauthoryear{Balbi, Bruni \& Quercellini}{Balbi et al.}{2007}]{b3} Balbi A., Bruni M., Quercellini C., 2007, Phys. Rev. D, 76, 103519
\bibitem[\protect\citeauthoryear{Barrow}{2004}]{b4} Barrow J. D., 2004, Class. Quant. Grav., 21, L79
\bibitem[\protect\citeauthoryear{Carneiro et al.}{2008}]{b5} Carneiro S., Dantas M. A., Pigozzo C., Alcaniz J. S., 2008, Phys. Rev. D, 77, 083504
\bibitem[\protect\citeauthoryear{Dabrowski}{2005}]{b6} Dabrowski M. P., 2005, Phys. Rev. D, 71, 103505
\bibitem[\protect\citeauthoryear{Dabrowski \& Denkiewicz}{2009}]{b7} Dabrowski M. P., Denkiewicz T., 2009, Phys. Rev. D, 79, 063521
\bibitem[\protect\citeauthoryear{Dabrowski \& Denkiewicz}{2010}]{b8} Dabrowski M. P., Denkiewicz T., 2010, in Alimi J. -M., F\"{u}zfa A., eds, Proc. AIP Conf. Ser. Vol. 1241, Invisible Universe. Am. Inst. Phys., New York, p. 561
\bibitem[\protect\citeauthoryear{Dabrowski, Denkiewicz \& Hendry}{Dabrowski et al.}{2007}]{b9} Dabrowski M. P., Denkiewicz T., Hendry M. A., 2007, Phys. Rev. D, 75, 123524 
\bibitem[\protect\citeauthoryear{Eisenstein \& Hu}{1998}]{b10} Eisenstein D. J., Hu W., 1998, ApJ, 496, 605 
\bibitem[\protect\citeauthoryear{Eisenstein et al.}{2005}]{b11} Eisenstein D. J. et al., 2005, ApJ, 633, 560
\bibitem[\protect\citeauthoryear{Elgaroy \& Multamaki}{2005}]{b12} Elgaroy O., Multamaki T., 2005, MNRAS, 356, 475
\bibitem[\protect\citeauthoryear{Elgaroy \& Multamaki}{2007}]{b13} Elgaroy O., Multamaki T., 2007, A\&A, 471, 65
\bibitem[\protect\citeauthoryear{Fernandez-Jambrina \& Lazkoz}{2004}]{b14} Fernandez-Jambrina L., Lazkoz R., 2004, Phys. Rev. D, 70, 121503
\bibitem[\protect\citeauthoryear{Hawking \& Ellis}{1973}]{b15} Hawking S. W., Ellis G. F. R., 1973, The Large Scale Structure of Space-Time, Cambridge Univ. Press, Cambridge
\bibitem[\protect\citeauthoryear{Kamenshchik, Kiefer \& Sandhofer}{Kamenshchik et al.}{2007}]{b16} Kamenshchik A., Kiefer C., Sandhofer B., 2007, Phys. Rev. D, 76, 064032
\bibitem[\protect\citeauthoryear{Koivisto}{2008}]{b17} Koivisto T., 2008, Phys. Rev. D., 77, 123513
\bibitem[\protect\citeauthoryear{Komatsu et al.}{2009}]{b18} Komatsu E. et al., 2009, ApJS, 180, 330
\bibitem[\protect\citeauthoryear{Komatsu et al.}{2011}]{b19} Komatsu E. et al., 2011, ApJS, 192, 18
\bibitem[\protect\citeauthoryear{Kowalski et al.}{2008}]{b20} Kowalski M. et al., 2008, ApJ, 686, 749
\bibitem[\protect\citeauthoryear{Krauss \& Chaboyer}{2003}]{b21} Krauss L. M., Chaboyer B., 2003, Science Mag., 299, 65
\bibitem[\protect\citeauthoryear{Nojiri \& Odintsov}{2004a}]{b22} Nojiri S., Odintsov S. D., 2004a, Phys. Lett. B, 595, 1
\bibitem[\protect\citeauthoryear{Nojiri \& Odintsov}{2004b}]{b23} Nojiri S., Odintsov S. D., 2004b, Phys. Rev. D, 70, 103522  
\bibitem[\protect\citeauthoryear{Nojiri, Odintsov \& Tsujikawa}{Nojiri et al.}{2005}]{b24} Nojiri S., Odintsov S. D., Tsujikawa S., 2005, Phys. Rev. D, 71, 063004
\bibitem[\protect\citeauthoryear{Riess et al.}{1998}]{b25} Riess A. G. et al., 1998, AJ, 116, 1009 
\bibitem[\protect\citeauthoryear{Riess et al.}{2009}]{b26} Riess A. G. et al., 2009, ApJ, 699, 539
\bibitem[\protect\citeauthoryear{Shafieloo et al.}{2006}]{b27} Shafieloo A., Alam U., Sahni V., Starobinsky A. A., 2006, MNRAS, 366, 1081
\bibitem[\protect\citeauthoryear{Shtanov \& Sahni}{2002}]{b28} Shtanov Y., Sahni V., 2002, Class. Quant. Grav., 19, L101
\bibitem[\protect\citeauthoryear{Spergel et al.}{2003}]{b29} Spergel D. N. et al., 2003, ApJS, 148, 175
\bibitem[\protect\citeauthoryear{Sola \& Stefancic}{2005}]{b30} Sola J., Stefancic H., 2005, Phys. Lett. B, 624, 147
\bibitem[\protect\citeauthoryear{Wang \& Mukherjee}{2007}]{b31} Wang Y., Mukherjee P., 2007, Phys. Rev. D, 76, 103533
 
\end{thebibliography}
\end{document}